\documentclass[pre,preprint,showkeys,amsfonts,amssymb,amsmath,eqsecnum,letterpaper,nofootinbib]{revtex4}

\usepackage{graphicx}

\begin{document}

\renewcommand{\r}{\mathbf{r}}
\newcommand{\x}{\mathbf{x}}
\newcommand{\y}{\mathbf{y}}
\newcommand{\F}[2]{\mathop{{}_#1F_#2}}
\newcommand{\simr}{\mathop{\underset{r\to 0}{\sim}}}
\newcommand{\eqat}[1]{\mathop{\underset{#1}{=}}}
\newcommand{\simat}[1]{\mathop{\underset{#1}{\sim}}}
\newcommand{\eqr}{\mathop{\underset{r\to 0}{=}}}
\newcommand{\I}{{\cal I}}
\newlength{\GraphicsWidth}
\setlength{\GraphicsWidth}{11cm}

\title{Short-distance expansion of correlation functions in the
  charge-symmetric two-dimensional two-component plasma: \\
  Exact results}

\author{Gabriel T\'ellez}
\email{gtellez@uniandes.edu.co} 
\affiliation{Departamento de F\'{\i}sica, Universidad de Los Andes,
A.A.~4976, Bogot\'a, Colombia}

\begin{abstract}
We determine exactly the short-distance leading behavior of the
density correlation functions of a two-dimensional two-component
charge-symmetric Coulomb gas composed of point particles, in the whole
regime of stability where the coulombic coupling $\beta<2$. More
generally, we compute the short-distance behavior of the effective
interaction potential between two external arbitrary charges $Q_1$ and
$Q_2$ immersed in the plasma, for $\beta |Q_1|<2$ and $\beta
|Q_2|<2$. We also find the short-distance asymptotics of the density
profiles near a single external charge $Q$ immersed in the plasma for
$\beta |Q|<2$. 
\end{abstract}



\maketitle


\section{Introduction}
\label{sec:Intro}

The two-dimensional charge-symmetric two-component plasma is a
two-dimensional Coulomb fluid composed of two species of point
particles with charges $\pm 1$. Each pair of particles (charges $q$
and $q'$) at a distance $r$ from each other interact via the
logarithmic Coulomb potential in two dimensions
\begin{equation}
  v(r)=-qq'\ln \frac{r}{L}
\end{equation}
where $L$ is an arbitrary length scale. We describe the system with
classical (i.e.~non quantum) statistical mechanics at a reduced
inverse temperature (coulombic coupling) $\beta$. Provided $\beta<2$
the system of point particles is stable against the collapse of
particles of opposite sign. We will consider here only the regime
$\beta<2$. To study the regime beyond $\beta=2$ it is necessary to
consider an additional repulsive short-range interaction to avoid the
collapse.

The two-component plasma is mappable into the quantum sine-Gordon
model~\cite{Samuel}, which in, two dimensions, is an integrable
model~\cite{Zamolod2,Destri-Vega,Zamolod-conformal-norm}. For
$\beta<2$, using results from the sine-Gordon model, the bulk
thermodynamic quantities of the two-component plasma have been
computed exactly~\cite{Samaj-Travenec-TCP}. Also, using recent results
for the form factors of the sine-Gordon
model~\cite{Lukyanov-form-fac1,Lukyanov-form-fac2}, the large-distance
behavior of the particle correlation functions have been determined
exactly~\cite{Samaj-Janco-TCP-large-distance-correl}. In
Ref.~\cite{Samaj-guest-charges}, the large-distance behavior of the
mean-force effective potential between two external charges immersed
in the plasma was determined. Also in that same
work~\cite{Samaj-guest-charges}, the large-distance behavior of the
density profiles around an external charge in the plasma was computed
(see also~\cite{Tellez-guest-charges}).

The work presented here is complementary to the ones of
Refs.~\cite{Samaj-Janco-TCP-large-distance-correl,Samaj-guest-charges,Tellez-guest-charges}.
We are interested here in the exact short-distance expansion of the
effective interaction energy between two external charges $Q_1$ and
$Q_2$ immersed in the plasma. We will extensively study the cases
$Q_2=\pm 1$ and $Q_1$ arbitrary and determine the short-distance
expansion of the density profile of around a single external charge in
the plasma. Also we will study the case $Q_1=1$ and $Q_2=\pm1$ which
allows us to determine the short-distance expansion of the density
correlation functions. Our main tool will be the operator product
expansion of the exponential fields of the sine-Gordon model.

This paper is organized as follows. In Sec.~\ref{sec:sine-Gordon} we
will recall some basics facts about the exact solution of the
two-component plasma and the sine-Gordon model needed for our
purposes. In Sec.~\ref{sec:OPE} follows a general discussion on the
short-distance expansion of the mean-force potential interaction
between two arbitrary external charges $Q_1$ and $Q_2$ immersed in the
plasma, which is obtained by means of the operator product
expansion. In Sec.~\ref{sec:charge-like}, we will consider the case of
charge-like particles $Q_1 Q_2 >0$ and in
Sec.~\ref{sec:charge-dislike} the case of particles with opposite
signs $Q_1 Q_2 <0$. Both sections~\ref{sec:charge-like}
and~\ref{sec:charge-dislike} have a similar organization, although
some technical details are different. First, we specialize in the case
where $|Q_1|=|Q_2|$ (identical or opposite external charges) and we
study the mean-force potential between these two external
charges. Then, we consider the case where $Q_2=\pm1$ is a charge of
the plasma and $Q_1=Q>0$ is arbitrary and we study the density profile
around the charge $Q$. Then, we consider the case $Q_2=\pm1$ and
$Q_1=1$ and determine the short-distance behavior of the density
correlation functions of the plasma. We test our general results valid
for arbitrary $\beta <2$ against the known results for the correlation
functions when $\beta=2$. In the appendices are presented some
expansions for particular values of $\beta$, $Q_1$ and $Q_2$ of the
expectation values of exponential fields and of some Coulomb-type
integrals (Dotsenko--Fateev
integrals~\cite{Dotsenko-Fateev-Coulomb-integrals1,Dotsenko-Fateev-Coulomb-integrals2})
that are needed in the main text.


\section{The two-component plasma and the sine-Gordon model}
\label{sec:sine-Gordon}

The statistical mechanics of the two-dimensional two-component plasma
are best worked out in the grand canonical ensemble with fugacity
$\zeta$ (with dimensions \textit{length}$^{-2}$). Writing down the
Boltzmann factor of the system one can notice that only the
combination $z=\zeta L^{\beta /2}$ is relevant. We shall still call
$z$ the fugacity, although it should be noticed that its dimensions
are \textit{length}$^{\beta/2-2}$. 

Doing a Hubbard-Stratonovich transformation, the grand canonical
partition function $\Xi$ of the plasma can be written as the partition
function of the quantum sine-Gordon model~\cite{Samuel}
\begin{equation}
  \Xi=\frac{\int {\cal D}\phi \exp[-S(z)]}{\int {\cal D}\phi \exp[-S(0)]}
\end{equation}
with 
\begin{equation}
  \label{eq:sG-action}
  S(z)=-\int d^2\r \left[
    \frac{1}{16\pi}\phi\Delta\phi+2z\cos(b\phi)
    \right]
\end{equation}
and
\begin{equation}
  b^2=\beta/4
  \,.
\end{equation}
Under this mapping, the bulk density and two-body densities of charges
$q=\pm1$ and $q'=\pm1$ are expressible
as~\cite{Samaj-Travenec-TCP,Samaj-Janco-TCP-large-distance-correl}
\begin{equation}
  \label{eq:bulk-density}
  n_q^b=z \langle e^{ibq\phi} \rangle
\end{equation}
and
\begin{equation}
  \label{eq:density-correl}
  n_{qq'}^{(2)}(|\r-\r'|)=z^2
  \langle e^{ibq\phi(\r)} e^{ibq'\phi(\r')} \rangle
\end{equation}
where the averages are taken with respect to the sine-Gordon
action~(\ref{eq:sG-action}).

In the sine-Gordon mapping the fugacity $z$ is renormalized by the
(diverging) term $e^{\beta
v(0)/2}$~\cite{Torres-Tellez-finite-size-DH}. Thus, to make a complete
connection between the sine-Gordon model and the statistical mechanics
problem of the two-component plasma it is necessary to specify the
proper renormalization scheme for the fugacity. In
Ref.~\cite{Zamolod-conformal-norm}, Zamolodchikov considered the
sine-Gordon model as the Gaussian conformal field (with action $S(0)$)
perturbed by the $\cos (b\phi)$ operator. Under this scheme, he used
the conformal normalization, where the exponential fields $e^{ib\phi}$
are normalized such that 
\begin{equation}
  \langle
  e^{ib\phi(\x_1)}\cdots  e^{ib\phi(\x_n)}
  e^{-ib\phi(\y_1)}\cdots e^{-ib\phi(\y_n)}
  \rangle_{z\to 0}=
  \frac{\prod_{1\leq j<i\leq n}
    |\x_i-\x_j|^{\beta}
    |\y_i-\y_j|^{\beta}}{\prod_{i,j=1}^{n} |\x_i-\y_j|^{\beta}}
\end{equation}
which is clearly the appropriate normalization for the Coulomb gas
system in view of relations~(\ref{eq:bulk-density})
and~(\ref{eq:density-correl}).

The specific dimensionless grand potential of the sine-Gordon model
was computed in Ref.~\cite{Destri-Vega} in terms of the mass of the
soliton of the sine-Gordon model. On the other hand the relationship
between the fugacity and the mass of the soliton under the conformal
normalization was found in Ref.~\cite{Zamolod-conformal-norm}. With
these results the density-fugacity relationship and full
thermodynamics of the Coulomb plasma can be
obtained~\cite{Samaj-Travenec-TCP}.  From the fugacity-density
relationship from Ref.~\cite{Samaj-Travenec-TCP} an explicit formula
for the expectation value of the exponential field, which we will need
in the following, can be obtained (see
also~\cite{Lukyanov-Zamolod-exp-field})
\begin{equation}
  \label{eq:eibphi}
  \langle e^{i b \phi} \rangle =2 
  \left(\frac{\pi z}{\gamma(\beta/4)}\right)^{\xi}
  \left(\frac{\Gamma(\xi/2)}{\Gamma(\frac{1+\xi}{2})}\right)^2
  \frac{\tan(\pi\xi/2)}{(4-\beta)\gamma(\beta/4)}
\end{equation}
where we use the notation $\gamma(x)=\Gamma(x)/\Gamma(1-x)$, with
$\Gamma(x)$ the Euler gamma function, and $\xi=\beta/(4-\beta)$.


\section{The interaction potential between two external charges and
  the operator product expansion} 
\label{sec:OPE}

Let us consider that two external point charges $Q_1$ and $Q_2$ are
immersed in the plasma and located at the origin and at $\r$
respectively. To avoid the collapse of the external charges with the
charges of the plasma of opposite sign, we suppose that $\beta
|Q_1|<2$ and $\beta |Q_2|<2$. We define the effective interaction
mean-force potential $E_{Q_1 Q_2}(r)$ between the charges $Q_1$ and
$Q_2$ as in~\cite{Samaj-guest-charges} by
\begin{equation}
  \exp[-\beta E_{Q_1 Q_2}(r)]=
  \frac{\Xi[Q_1,0;Q_2,\r]/\Xi}{(\Xi[Q_1]/\Xi)(\Xi[Q_2]/\Xi)}
\end{equation}
where $\Xi[Q_1,0;Q_2,\r]$ is the grand partition function of the
plasma in the presence of the charge $Q_1$ at the origin and of the
charge $Q_2$ at $\r$ (with the direct interaction $-Q_1 Q_2 \ln r$
between the external charges included in the Hamiltonian) and $\Xi[Q]$
is the grand partition function of the plasma in the presence of a
single external charge $Q$. As shown in
Ref.~\cite{Samaj-guest-charges}, the effective interaction can be
expressed in terms of expectation values of exponentials fields of the
sine-Gordon model
\begin{equation}
  \exp[-\beta E_{Q_1 Q_2}(r)]=
  \frac{\langle e^{i b Q_1 \phi(0)} e^{i b Q_2 \phi(\r)} \rangle
  }{\langle e^{i b Q_1 \phi} \rangle\langle e^{i b Q_2 \phi} \rangle}
  \,.
\end{equation}

The denominator in the preceding expression can be computed since an
exact expression for the expectation value $\langle e^{i b Q \phi}
\rangle$ was conjectured in Ref.~\cite{Lukyanov-Zamolod-exp-field}, a
conjecture later supported by several other
studies~\cite{Fateev-reflexion,
Bajnok-numerical-exp-field,Lu-variational-exp-field,Samaj-guest-charges},
\begin{equation}
  \label{eq:eibQ}
  \langle e^{i b Q \phi} \rangle
  = \left(\frac{\pi z}{\gamma(\beta/4)}\right)^{Q^2 \xi} \exp[I_b(Q)]
\end{equation}
with
\begin{equation}
  I_b(Q)=\int_0^\infty
  \frac{dt}{t}
  \left[
    \frac{\sinh^2(2Qb^2 t)}{2\sinh(b^2 t)\sinh(t)\cosh[(1-b^2)t]}
    -2Q^2 b^2 e^{-2t}
    \right]
  \,.
\end{equation}
This expression is valid for $\beta|Q|<2$, otherwise the integral
$I_b(Q)$ diverges. Notice that this is the same stability condition
necessary to avoid the collapse of an external charge $Q$ with the
counterions of the plasma (ions with charge of opposite sign than
$Q$).
  
On the other hand, the short-distance expansion of $\langle e^{i b Q_1
\phi(0)} e^{i b Q_2 \phi(\r)} \rangle$ can be obtained by means of the
operator product expansion~\cite{Wilson-OPE}, which
reads~\cite{Fateev-Fradkin-Lukyanov-Zad2-descendent}
\begin{equation}
  \label{eq:OPE}
  e^{i b Q_1 \phi(0)} e^{i b Q_2 \phi(\r)}=
  \sum_{n=-\infty}^{n=+\infty}
  C_{Q_1 Q_2}^{n,0}(r) e^{ib (Q_1+Q_2+n)\phi}+\cdots
\end{equation}
where the dots denote subdominant contributions from descendant fields
(like $(\partial\phi)^2(\bar{\partial}\phi)^2\, e^{i(Q_1+Q_2+n)\phi}$ and so
on), which will not be needed for our purposes. The coefficients of
the operator product expansion have the following
form~\cite{Fateev-Fradkin-Lukyanov-Zad2-descendent}
\begin{equation}
    \label{eq:OPE-coefs-C}
  C_{Q_1 Q_2}^{n,0}(r)=z^{|n|}
  r^{\beta Q_1 Q_2 + n \beta (Q_1+Q_2) + 2|n|(1-\frac{\beta}{4}) +n^2
    \beta /2} f_{Q_1 Q_2}^{n,0}(z^2 r^{4-\beta})
\end{equation}
where $f_{Q_1 Q_2}^{n,0}$ admit a power series expansion of the form
\begin{equation}
  \label{eq:OPE-f}
  f_{Q_1 Q_2}^{n,0}(x)=\sum_{k=0}^{\infty} f_{k}^{n,0}(Q_1,Q_2) x^k
  \,.
\end{equation}
The coefficients of the first leading terms of the
expansion~(\ref{eq:OPE-f}) are
\begin{eqnarray}
  \label{eq:OPE-coefs-f}
  f_{0}^{0,0}(Q_1,Q_2)&=&1\\
  f_{0}^{n,0}(Q_1,Q_2)&=&j_n(\beta Q_1,\beta Q_2,
  \beta)\qquad\text{for\ }n>0\\
  f_{0}^{n,0}(Q_1,Q_2)&=&j_{|n|}(-\beta Q_1,-\beta Q_2,
  \beta)\qquad\text{for\ }n<0
\end{eqnarray}
where
\begin{equation}
  \label{eq:Dotsenko-integrals-j-def}
  j_n(\beta Q_1,\beta Q_2,\beta) =\frac{1}{n!}\int \prod_{k=1}^n d^2
  x_k \prod_{k=1}^n |x_k|^{\beta Q_1} |1-x_k|^{\beta Q_2} 
  \prod_{1\leq k<j\leq n}
  |x_k-x_j|^{\beta} \,.
\end{equation}
These Coulomb-type integrals have been computed exactly by Dotsenko
and Fateev~\cite{Dotsenko-Fateev-Coulomb-integrals1,
Dotsenko-Fateev-Coulomb-integrals2}
\begin{eqnarray}
  \label{eq:Dotsenko-integrals-j}
  j_n(\beta Q_1,\beta Q_2,\beta)&=&
  \pi^{n} \gamma(\beta/4)^{-n} \prod_{k=1}^n \gamma (k\beta/4)
  \times
  \\
  && \hspace{-3cm}
  \prod_{k=0}^{n-1}
  \gamma\left(1+\frac{\beta Q_1}{2}+k\frac{\beta}{4}\right)
  \gamma\left(1+\frac{\beta Q_2}{2}+k\frac{\beta}{4}\right)
  \gamma\left(-1-\frac{\beta}{2}
  (Q_1+Q_2)-(n-1+k)\frac{\beta}{4}\right)
  \,.
  \nonumber
\end{eqnarray}
The coefficient for $n=0$ of the first subleading term ($k=1$)
in~(\ref{eq:OPE-f}) reads
\begin{equation}
  f_{1}^{0,0}(Q_1,Q_2)=J(\beta Q_1,\beta Q_2,\beta)
\end{equation}
where
\begin{equation}
  \label{eq:Dotsenko-integral-J}
  J(\beta Q_1,\beta Q_2,\beta)=
  \int d^2 x \,d^2 y\,
  \frac{|x|^{\beta Q_1} |1-x|^{\beta Q_2}}{|y|^{\beta Q_1}
  |1-y|^{\beta Q_2} |x-y|^{\beta}}
\end{equation}
which can be expressed in terms of hypergeometric functions $\F{3}{2}$
at unity~\cite{Dotsenko-Picco-Pujol,Guida-Magnoli}. In
Sec.~\ref{sec:charge-dislike} and in the appendix~\ref{app:J} we
will compute a special case of this integral when $Q_1=-Q_2$.

Putting all these expressions together, the first terms of the
short-distance expansion of the Boltzmann factor of $E_{Q_1 Q_2}(r)$
are
\begin{eqnarray}
  \label{eq:small-r-general}
  &&\exp[-\beta E_{Q_1 Q_2}(r)]= \frac{\langle e^{i b (Q_1+Q_2) \phi}
    \rangle}{ \langle e^{i b Q_1\phi} \rangle \langle e^{i b Q_2\phi}
    \rangle } \,r^{\beta Q_1 Q_2} \\ 
  &&+ \sum_{n=1}^{\infty}
  \frac{\langle e^{i b (Q_1+Q_2+n) \phi} \rangle}{ \langle e^{i b
      Q_1\phi} \rangle \langle e^{i b Q_2\phi} \rangle } j_n(\beta
  Q_1,\beta Q_2,\beta)\, z^{n} r^{\beta Q_1 Q_2 + n \beta (Q_1+Q_2)
    + 2 n(1-\frac{\beta}{4}) +n^2 \beta /2}\{1+O(z^2 r^{4-\beta})\}
  \nonumber\\ 
  &&+ \sum_{n=-\infty}^{-1} \frac{\langle e^{i b
      (Q_1+Q_2+n) \phi} \rangle}{ \langle e^{i b Q_1\phi} \rangle
    \langle e^{i b Q_2\phi} \rangle } j_{|n|}(-\beta Q_1,-\beta
  Q_2,\beta)\, z^{|n|} r^{\beta Q_1 Q_2 + n \beta (Q_1+Q_2) +
    2|n|(1-\frac{\beta}{4}) +n^2 \beta /2}\{1+O(z^2 r^{4-\beta})\}
  \nonumber\\ 
  &&+ \frac{\langle e^{i b (Q_1+Q_2) \phi} \rangle}{
    \langle e^{i b Q_1\phi} \rangle \langle e^{i b Q_2\phi} \rangle }
  \,z^2 r^{\beta Q_1 Q_2+4-\beta} J(\beta Q_1,\beta Q_2,\beta) \{ 1
  + O(z^2 r^{4-\beta})\}+O(r^{\beta Q_1 Q_2 + 4}) 
  \nonumber
  \,.
\end{eqnarray}
The $O(r^{\beta Q_1 Q_2 + 4})$ term in the last line comes from a
descendant field contribution which is explicitly computed in
Ref.~\cite{Fateev-Fradkin-Lukyanov-Zad2-descendent}, but that will not
be needed in the following.

If $\beta Q_1$, $\beta Q_2$ and $\beta$ are small enough the dominant
contribution is given by the term $n=0$ in~(\ref{eq:OPE})
\begin{equation}
  \label{eq:small-r-dominant-n=0}
  \exp[-\beta E_{Q_1 Q_2}(r)]\simr
  \frac{\langle e^{i b (Q_1+Q_2) \phi}
    \rangle}{ \langle e^{i b Q_1\phi} \rangle \langle e^{i b Q_2\phi}
    \rangle } \,r^{\beta Q_1 Q_2}\,
\end{equation}
which is an expected result: the two charges interact at short
distances with the bare Coulomb potential $-Q_1 Q_2 \ln r$. Notice
however that depending on the values of $\beta Q_1$, $\beta Q_2$ or
$\beta$, some of the following terms in the
expansion~(\ref{eq:small-r-general}) can become dominant over the
$n=0$ term. We will explore this possibility in the following sections.


\section{The charge-like case: $Q_1>0$ and $Q_2>0$}
\label{sec:charge-like}

Let us consider that $Q_1$ and $Q_2$ have the same sign.  In the
general small-$r$ asymptotics~(\ref{eq:small-r-general}), the $n=-1$
term will be dominant over the $n=0$ term when $\beta(Q_1+Q_2) > 2$, a
condition which will take place if $Q_1+Q_2$ is positive and
sufficiently large. On the other hand the term $n=1$ will be dominant
over the $n=0$ term if $\beta(Q_1+Q_2) < -2$. As we see, depending on
the sign of $Q_1+Q_2$, the terms of the series for $n<0$ (if
$Q_1+Q_2>0$) or $n>0$ (if $Q_1+Q_2<0$) will become dominant. 

Let us suppose, without loss of generality, that $Q_1>0$ and
$Q_2>0$. The term $-n$ ($n> 0$) will be the dominant term of the
series~(\ref{eq:small-r-general}) if
$2+(n-1)\beta<\beta(Q_1+Q_2)<2+n\beta$. Thus, the dominant term of the
small-$r$ expansion of $\exp[-\beta E_{Q_1 Q_2}(r)]$ is
\begin{equation}
  \label{eq:small-r-general-dominant}
  e^{-\beta E_{Q_1 Q_2}(r)}
  \simr
  \begin{cases}
      \frac{\langle e^{i b (Q_1+Q_2) \phi}
    \rangle}{ \langle e^{i b Q_1\phi} \rangle \langle e^{i b Q_2\phi}
    \rangle } \,r^{\beta Q_1 Q_2}\,,
      &  \hspace{-5cm}
      0<\beta(Q_1+Q_2)<2\\
  \frac{\langle e^{i b
      (Q_1+Q_2-n) \phi} \rangle}{ \langle e^{i b Q_1\phi} \rangle
    \langle e^{i b Q_2\phi} \rangle } j_{n}(-\beta Q_1,-\beta
  Q_2,\beta)\, z^{n} r^{\beta Q_1 Q_2 - n \beta (Q_1+Q_2) +
    2n(1-\frac{\beta}{4}) +n^2 \beta /2}
  \,,
    \\&
  \hspace{-10cm}
  2+(n-1)\beta < \beta (Q_1+Q_2) <2+n\beta;
  \ \ n=1,2,\ldots
  \end{cases}
\end{equation}
This relation is valid provided we are still in the region of
stability of the system, thus $\beta Q_1<2$, $\beta Q_2<2$ and
$\beta<2$.

There is an interesting relation between the changes of the dominant
small-$r$ behavior of the effective potential shown
in~(\ref{eq:small-r-general-dominant}) and the corresponding
coefficient expressed in terms of the Dotsenko-Fateev
integrals~(\ref{eq:Dotsenko-integrals-j-def}).  $j_n(-\beta Q_1,-\beta
Q_2, \beta)$ expressed as an integral in
equation~(\ref{eq:Dotsenko-integrals-j-def}) can be interpreted as the
partition function of a Coulomb gas with the external charges $Q_1$ at
0 and $Q_2$ at 1 and $n$ charges $-1$. Thus, the behavior of $E_{Q_1
Q_2}(r)$ when $2+(n-1)\beta < \beta (Q_1+Q_2) <2+n\beta$ is dominated
by the approach of $n$ charges $-1$ to the positive external charges
$Q_1$ and $Q_2$. This confirms and generalizes the analysis of Hansen
and Viot~\cite{Hansen-Viot}.

The integral~(\ref{eq:Dotsenko-integrals-j-def}) defining $j_n(-\beta
Q_1,-\beta Q_2, \beta)$ converges at $x_k\to 0$ if $\beta Q_1<2$, at
$x_k\to 1$ if $\beta Q_2<2$ and at $|x_k|\to \infty$ if
$\beta(Q_1+Q_2)>2 +(n-1)\beta$. On the other hand, $\langle e^{i
b(Q_1+Q_2-n) \phi} \rangle$, converges if
$\beta(Q_1+Q_2)<2+n\beta$. These are precisely the same conditions for
the corresponding term in the small-$r$
expansion~(\ref{eq:small-r-general}) to become dominant. Thus when the
term $-n$ of the general operator product
expansion~(\ref{eq:small-r-general}) becomes dominant, its
corresponding coefficient in~(\ref{eq:small-r-general-dominant}) is
properly defined.

When the condition $\beta(Q_1+Q_2)>2 +(n-1)\beta$ is not satisfied,
the integral defining $j_n$ in
equation~(\ref{eq:Dotsenko-integrals-j-def}) is not convergent, but
the corresponding term in~(\ref{eq:small-r-general}) is not the
dominant one. To evaluate this subdominant contribution one has to
resort to the analytic continuation of $j_n$ provided by the
expression~(\ref{eq:Dotsenko-integrals-j}) in terms of Gamma
functions.

At $\beta(Q_1+Q_2)=2+(n-1)\beta$, the term $-n$ and the term $-n+1$
of~(\ref{eq:small-r-general}) are of the same order, in $r^{\beta Q_1
Q_2-n(n-1)\beta/2}$, and they should be added to obtain the correct
leading behavior. The coefficient of the term $-n$ exhibits a pole of
first order due to $j_n(-\beta Q_1,-\beta Q_2,\beta)$, but the
coefficient of the term $-n+1$ of the
expansion~(\ref{eq:small-r-general}) also exhibits a pole of first
order due to $ \langle e^{i b(Q_1+Q_2-n+1) \phi} \rangle$.  Indeed, if
we write $\beta(Q_1+Q_2)=2+(n-1)\beta+\beta \epsilon$, we
have~\cite{Samaj-guest-charges}
\begin{equation}
  \langle
  e^{ib(Q_1+Q_2-n+1)}
  \rangle
  \simat{\epsilon\to0}
  - \langle 
  e^{ib(\frac{1}{2b^2}-1)}
  \rangle
  \frac{2\pi z}{\beta \epsilon}
\end{equation}
and
\begin{eqnarray}
  j_n(-\beta Q_1, -\beta Q_2,\beta)&=&\pi
  j_{n-1}(-\beta Q_1, -\beta Q_2,\beta)
  \frac{\gamma(\frac{n\beta}{4})\gamma(\frac{\beta}{4}
    +\frac{\beta\epsilon}{2})}{
    \gamma(\frac{n\beta}{4}+\frac{\beta\epsilon}{2})}
  \times
  \nonumber\\
  &&\gamma(1-\beta Q_1+\frac{(n-1)\beta}{4})
  \gamma(\beta Q_1+\frac{(n-1)\beta}{4}-\frac{\beta\epsilon}{2})
  \gamma(\frac{\beta\epsilon}{2})
  \nonumber\\
  &\simat{\epsilon\to0}&
  j_{n-1}(-\beta Q_1, -\beta Q_2,\beta)\frac{2\pi}{\beta \epsilon}
  \,.
\end{eqnarray}
Both coefficients turn out to have the same residue in
absolute value but with opposite signs, and thus cancel each other,
giving a finite value, and an additional $r^{\beta Q_1
Q_2-n(n-1)\beta/2}\ln r$ contribution to the small-$r$ expansion. This
will be illustrated in several particular cases in the following
sections.


\subsection{Effective mean-force potential between two identical
  external charges}
\label{sec:+Q+Q}

In this section we consider two identical external charges immersed in
the plasma $Q_1=Q_2=Q>0$. Using
equation~(\ref{eq:small-r-general-dominant}), the first few changes in
the small-$r$ asymptotics of the effective potential read
\begin{equation}
  e^{-\beta E_{Q Q}(r)}\simr
  \begin{cases}
    \displaystyle\frac{\langle e^{2 i b
       Q \phi} \rangle}{ \langle e^{i b Q\phi} \rangle^2 }
    \,r^{\beta Q^2}&0<\beta Q<1\\
    \displaystyle\frac{\langle e^{i b
      (2 Q-1) \phi} \rangle}{ \langle e^{i b Q\phi} \rangle^2 } 
    j_{1}(-\beta Q,-\beta Q,\beta)\, 
    z \,r^{\beta Q^2 - 2 (\beta Q-1)} & 1<\beta Q<1+\beta/2\\
    \displaystyle\frac{\langle e^{i b
      (2 Q-2) \phi} \rangle}{ \langle e^{i b Q\phi} \rangle^2 } 
  j_{2}(-\beta Q,-\beta Q,\beta)\, 
  z^{2} r^{\beta Q^2 - 4(\beta Q-1)+\beta}&
    1+\frac{\beta}{2}<\beta Q< 1 + \beta\,; \beta <1\\
  \frac{\langle e^{i b
      (2 Q-n) \phi} \rangle}{ \langle e^{i b Q\phi} \rangle^2 } 
  j_{n}(-\beta Q,-\beta Q,\beta)\, 
  z^{n} r^{\beta Q^2 - 2 n \beta Q +
    2n(1-\frac{\beta}{4}) +n^2 \beta /2}  
  \\ 
  &
  \hspace{-4cm}
 1+\frac{(n-1)\beta}{2} < \beta Q <1 +\frac{n\beta}{2}\,
  ; \beta<\frac{2}{n-1}\,;
   n=3, 4, \ldots
  \end{cases}
\end{equation}
Notice that in order to satisfy the relation $\beta Q >
1+(n-1)\beta/2$ and the stability condition $\beta Q<2$, the coulombic
coupling must be small enough: $\beta <2/(n-1)$. Thus, as $\beta$
becomes larger there are few changes in the small-$r$ behavior of the
effective potential.

\subsection{Coion density profile around an single external charge
  $Q>0$}
\label{sec:coion}

Let us consider now a single external charge $Q>0$ at the origin
immersed in the plasma. The density of the particles with charge $+1$
(coions) of the plasma is given by~\cite{Samaj-guest-charges,
Tellez-guest-charges}
\begin{equation}
  n_{+}(r)=z \frac{ \langle e^{i b Q\phi(0)} e^{i b \phi(\r)}
  \rangle}{\langle e^{i b Q\phi}\rangle}
\end{equation}
or equivalently, using relation~(\ref{eq:bulk-density}),
\begin{equation}
n_{+}(r)=n_{+}^b \frac{ \langle e^{i b Q\phi(0)} e^{i b \phi(\r)}
  \rangle}{\langle e^{i b Q\phi}\rangle \langle e^{i b\phi}\rangle}
\end{equation}
Thus, using the expansion~(\ref{eq:small-r-general}) with $Q_1=Q$ and
$Q_2=+1$ we can deduce the small-$r$ asymptotics of $n_{+}(r)$.

To ensure the stability of the system we require that $\beta
Q<2$. This condition reduces the number of changes in the small-$r$
behavior~(\ref{eq:small-r-general-dominant}) to two. Writing
explicitly the $j_1$ coefficient we find that the dominant behavior of
the coion density is
\begin{equation}
  \label{eq:n+}
  n_{+}(r)\simr
  \begin{cases}
    z\, \frac{\langle e^{ib (Q+1) \phi} \rangle}{\langle e^{ib Q\phi}\rangle}
    \,
    r^{\beta Q}\,, &
    0<\beta Q<2-\beta \\
    \pi z^2 \gamma\left(1-\frac{\beta Q}{2}\right) 
    \gamma\left(1-\frac{\beta}{2}\right)
    \gamma\left(-1+\frac{\beta(Q+1)}{2}\right)\,
    r^{2-\beta}\,,
    & 2-\beta<\beta Q<2
  \end{cases}
\end{equation}
Where $\langle e^{ib (Q+1) \phi} \rangle$ and $\langle e^{ibQ \phi}
\rangle$ can be computed from equation~(\ref{eq:eibQ}). This confirms
and completes the analysis of Ref.~\cite{Tellez-guest-charges}, where
the change of behavior at $\beta Q=2-\beta$ of the coion density was
identified and interpreted as a ``precursor'' of the counterion
condensation, in the sense that for $\beta Q>2-\beta$, the effective
potential (times $\beta$) of a coion and the charge $Q$ behaves at
short distances as the bare Coulomb potential with a reduced charge:
$-(2-\beta)\ln r$, instead of $-\beta Q \ln r$. The charge is reduced
because of the counterion condensation. With the present work, we have
also an explicit expression for the coefficient multiplying
$r^{2-\beta}$.

It is also interesting to study the transition at $\beta Q=2-\beta$
and at $\beta Q =2$. At $\beta Q=2-\beta$, both terms $r^{\beta Q}$
and $r^{2-\beta}$ shown in equation~(\ref{eq:n+}) become of the same
order and should be added to obtain the correct dominant behavior.

Using the expansions of the expectation values of exponentials
field~(\ref{eq:eibQ+eps}) and~(\ref{eq:eibQ-betaQ=2}) computed in the
appendix~\ref{app:eibQphi} we have
\begin{multline}
  \label{eq:eibQ+1/eiQ}
  \frac{\langle e^{ib (Q+1) \phi} \rangle}{\langle e^{ib Q \phi}
  \rangle}
  \eqat{\beta Q\to2-\beta}
  \frac{2\pi z}{2-\beta(Q+1)}
  \times
  \\
  \left\{
    1+\frac{2-\beta(Q+1)}{\beta}\left[R_1(b^2)+I'_b(\frac{1}{2b^2}-1)
      -\frac{2\beta}{4-\beta}\ln\left(\frac{\pi z}{\gamma(b^2)}\right)
      \right]
    + O([\beta(Q+1)-2]^2)
    \right\}
\end{multline}
where $I_b'(Q)=\partial I_b(Q)\partial Q$ and $R_1(b^2)$ are given in
equations~(\ref{eq:Iprimeb}) and~(\ref{eq:R1}) of the
appendix~\ref{app:eibQphi}). On the other hand the
$\gamma(-1+\beta(Q+1)/2)$ from $j_1$ also exhibit a pole at $\beta
Q=2-\beta$, but with opposite residue:
\begin{multline}
  \label{eq:j1-bQ=2-beta}
  \pi z \gamma\left(1-\frac{\beta Q}{2}\right) 
    \gamma\left(1-\frac{\beta}{2}\right)
    \gamma\left(-1+\frac{\beta(Q+1)}{2}\right)
    \eqat{\beta Q\to 2-\beta}\\
    \frac{2\pi z}{2-\beta(Q+1)}
    \left[
      -1 - \frac{2-\beta(Q+1)}{2}\left[2C+
      \psi(1-\beta/2)+\psi(\beta/2)\right]
      +O([\beta(Q+1)-2]^2)
      \right]
\end{multline}
where $\psi(x)=\Gamma'(x)/\Gamma(x)$ is the digamma function and
$C=-\psi(1)$ is the Euler constant. Thus adding both contribution
gives a finite result. Using~(\ref{eq:eibQ+1/eiQ}),
(\ref{eq:j1-bQ=2-beta}) and expanding $r^{\beta Q}
=r^{2-\beta}(1+[\beta(Q+1)-2]\ln r+O([\beta(Q+1)-2]^2)$ finally
yields, for $\beta Q =2-\beta$,
\begin{multline}
  \label{eq:n+betaQ=2-beta}
  n_{+}(r)\simr
  2\pi z^2 r^{2-\beta}
  \times\\
  \left\{
  -\ln\left[ \left(\frac{\pi
  z}{\gamma(b^2)}\right)^{\frac{2}{4-\beta}}
    r\right]
  +\frac{1}{\beta}\left[R_1(b^2)+I_b'(\frac{1}{2b^2}-1)\right]
  -C -\frac{1}{2}\left[ \psi(1-\beta/2)+\psi(\beta/2)\right]
  \right\}
\end{multline}
Notice that the power of $z$ in the logarithm, $z^{2/(4-\beta)}$, is
the appropriate one to have a dimensionless argument in the logarithm.

When $\beta Q=2$, the contribution from the term $n=-2$
of~(\ref{eq:small-r-general}) in $r^{-\beta Q-\beta+4}$, is of the
same order as the one for $n=-1$, in $r^{2-\beta}$. The coefficients
of both terms exhibit a pole with opposite residues when $\beta
Q=2$. A similar mechanism as before applies, adding both contributions
gives a finite result. The term $n=-2$ of the expansion of $n_{+}(r)$
reads
\begin{equation}
  z^3 r^{-\beta Q-\beta+4}
  \frac{\langle e^{ib(Q-1)\phi}\rangle}{\langle
  e^{ibQ\phi}\rangle}
  j_2(-\beta Q,-\beta,\beta)
\end{equation}
with
\begin{equation}
  j_2(-\beta Q,-\beta,\beta)=
  \pi^2 \gamma(1-\frac{\beta}{4})^2\gamma(1-\frac{\beta Q}{2})
  \gamma(-1+\frac{\beta Q}{2})
  \gamma(-1+\frac{\beta Q}{2}+\frac{\beta}{4})
  \gamma(1-\frac{\beta Q}{2}+\frac{\beta}{4})
\end{equation}
When $\beta Q\to 2$, we have
\begin{multline}
  \label{eq:n+-betaQ=2-contrib1}
\frac{\langle e^{ib(Q-1)\phi}\rangle}{\langle
  e^{ibQ\phi}\rangle}
  j_2(-\beta Q,-\beta,\beta)
  \eqat{\beta Q\to 2}
  \frac{2\pi}{z(\beta Q-2)}
  \times\\
  \left\{
    1+\frac{\beta Q-2}{\beta}
    \left[
      R_1(b^2)+I'_b(\frac{1}{2b^2}-1)-2\xi\ln\frac{\pi z}{\gamma(b^2)}
      \right]
    +O((\beta Q-2)^2)
    \right\}
\end{multline}
while from the term $n=-1$ we have a contribution from the $j_1$
coefficient:
\begin{equation}
  \label{eq:n+-betaQ=2-contrib2}
  \gamma\left(1-\frac{\beta Q}{2}\right) 
    \gamma\left(1-\frac{\beta}{2}\right)
    \gamma\left(-1+\frac{\beta(Q+1)}{2}\right)
    \eqat{\beta Q\to 2}
    \frac{2}{2-\beta Q} -2C -\psi(1-\beta/2)-\psi(\beta/2)
\,.
\end{equation}
Expanding $r^{-\beta Q-\beta+4}=r^{2-\beta}[1+(2-\beta Q)\ln r
+O((2-\beta Q)^2)]$, and adding all
contributions~(\ref{eq:n+-betaQ=2-contrib1})
and~(\ref{eq:n+-betaQ=2-contrib2}), we actually find that at $\beta
Q=2$, the small-$r$ behavior of $n_{+}(r)$ is exactly the same as the
one for $\beta Q=2-\beta$ given by equation~(\ref{eq:n+betaQ=2-beta}).


\subsection{Density correlation function $n_{++}^{(2)}(r)$}
\label{sec:n++}

The two-body density correlation function between particles of the
same sign (say positive), can be expressed as
\begin{equation}
  n_{++}^{(2)}(r)=z^2 \langle e^{ib\phi(0)} e^{ib\phi(\r)}
  \rangle
  =(n_{+}^b)^2
  \,
  \frac{\langle e^{ib\phi(0)} e^{ib\phi(\r)}\rangle}{
    \langle e^{ib\phi}\rangle^2}
  \,.
\end{equation}
Thus, using the results of the last section with $Q=1$ allow us to find
the small-$r$ behavior of the correlation function:
\begin{equation}
  \label{eq:n++}
  n_{++}^{(2)}(\r)\simr
  \begin{cases}
    z^2 \langle e^{2ib\phi} \rangle\,r^{\beta}
    & \beta <1\\
    2\pi z^2 n_{+}^b
    \,r
    \left\{
    -\ln\left[\left(\frac{\pi z}{\gamma(1/4)}\right)^{2/3}\,r\right]+
    R_1(1/4)+I'_{1/2}(1)+2\ln 2
    \right\}
    & \beta=1 \\
    \pi z^2 n_{+}^b \gamma(1-\frac{\beta}{2})^2
    \gamma(\beta-1)\, r^{2-\beta}=
    \pi z^3 \langle e^{ib\phi} \rangle \gamma(1-\frac{\beta}{2})^2
    \gamma(\beta-1)\, r^{2-\beta} & 1<\beta<2
  \end{cases}
\end{equation}
A numerical estimate of the constant intervening in the case $\beta=1$
is $R_1(1/4)+I'_{1/2}(1)\simeq-1.73246$.

Our result~(\ref{eq:n++}) confirms the change of behavior of the
correlation function $n_{++}^{(2)}(r)$ at $\beta=1$, predicted by
Hansen and Viot~\cite{Hansen-Viot}. Furthermore, we obtained an
explicit analytical expression for the coefficients multipliying
$r^{\beta}$ (for $\beta<1$) and $r^{2-\beta}$ (for $\beta>1$).

The limit $\beta\to 2$ cannot be obtained from the results of the
previous section because there is an additional divergence due to
$n_{+}^b$. When $\beta\to 2$, the term $n=-1$ of the general
series~(\ref{eq:small-r-general}), in $r^{2-\beta}$ is of the same
order as the term $n=-2$. This term ($n=-2$) gives a contribution to
$n_{++}^{(2)}(r)$ equal to
\begin{equation}
  \label{eq:n++n=-2}
  \pi^2 z^4 r^{4-2\beta} \gamma(1-\frac{\beta}{4})^3
  \gamma(1-\frac{\beta}{2}) \gamma(\frac{\beta}{2}-1)
  \gamma(-1+\frac{3\beta}{4})
  =
   -\frac{4 \pi^2 z^4 r^{4-2\beta}}{(2-\beta)^2}
    \gamma(1-\frac{\beta}{4})^3
  \gamma(-1+\frac{3\beta}{4})
  \,.
\end{equation}
Notice that the coefficients of $r^{2-\beta}$ in~(\ref{eq:n++}) and
of $r^{4-2\beta}$ in~(\ref{eq:n++n=-2}) exhibit now a pole of second
order when $\beta\to 2$, as opposed to the situation of the previous
section, when the pole was of order one. Indeed,
\begin{equation}
  \label{eq:n++contrib1}
  \frac{-4\gamma(1-\frac{\beta}{4})^3
  \gamma(-1+\frac{3\beta}{4})}{(\beta-2)^2}
  \eqat{\beta\to2}
  -\frac{4}{(\beta-2)^2}+O(\beta-2)
\end{equation}
while, using the expansion~(\ref{eq:eibphi-beta=2}) of $\langle
e^{ib\phi} \rangle$, and expanding the gamma functions,
\begin{multline}
  \label{eq:n++contrib2}
  \langle e^{ib\phi} \rangle \gamma(1-\frac{\beta}{2})^2
    \gamma(\beta-1)
    \eqat{\beta\to2}
    \frac{8\pi z}{(\beta-2)^2}
    \Bigg[
      1+(\beta-2)[C+\ln(\pi z)]\\
      +\frac{(\beta-2)^2}{2}[C+\ln(\pi z)]
      [1+C+\ln(\pi z)] 
      \Bigg]+O(\beta-2)
\end{multline}
Notice that now the poles do not cancel each other. Expanding
$r^{2-\beta}=1+(2-\beta)\ln r +\frac{(2-\beta)^2}{2}(\ln
r)^2+O((\beta-2)^3)$ and $r^{4-2\beta}=1+2(2-\beta)\ln r
+2(2-\beta)^2(\ln r)^2+O((\beta-2)^3)$, and putting together all
contributions from~(\ref{eq:n++contrib1}) and~(\ref{eq:n++contrib2}),
we finally find
\begin{multline}
  n^{(2)}_{++}(r)
  \simat{r\to0,\,\beta\to2}
  (2\pi z^2)^2
  \Bigg[
    \frac{1}{(\beta-2)^2}
    + \frac{2}{\beta-2}\, [C+\ln(\pi z)]
    \\
    -[\ln(\pi z r)+C]^2
    +2[C+\ln(\pi z)]^2+C+\ln(\pi z)
    +O(\beta-2)
    \Bigg]
\end{multline}
In the preceding expression we recognize the
expansion~(\ref{eq:eibphi-beta=2-carre}) of $\langle e^{ib\phi}
\rangle^2$ when $\beta\to 2$. Indeed, we have
\begin{equation}
  n_{++}^{(2)}(r)
  \simat{r\to0,\,\beta\to2}
  -(2\pi z^2)^2 \left(\ln(\pi z r)+C\right)^2
  +z^{2} \langle e^{ib\phi}\rangle^2
\end{equation}
Remembering that $n_{+}^{b}=z \langle e^{ib\phi}\rangle$, we recover a
known result at $\beta=2$. The correlation function $n_{++}^{(2)}(r)$
does not have a finite limit when $\beta=2$, however the truncated
correlation function $n_{++}^{(2)T} (r)=n_{++}^{(2)}(r) -(n_{+}^b)^2$,
does have a finite limit at the collapse point $\beta=2$
\begin{equation}
  \label{n++-beta=2-asympt}
  n_{++}^{(2)T}(r)
  \simat{r\to0}
  -(2\pi z^2)^2 \left(\ln(\pi z r)+C\right)^2
  \,, \qquad \beta=2
\end{equation}
Furthermore, the density correlation functions of the two-component
plasma at $\beta=2$ have been computed exactly in
Refs.~\cite{CornuJanco-Coulomb-gas, CornuJanco-JCP}
\begin{equation}
  \label{eq:n++-beta=2-exact}
  n_{++}^{(2)T}(r)=-(2\pi z^2)^2
  \left(K_0(2\pi z r)\right)^2
\end{equation}
where $K_0$ is the modified Bessel function of order 0. Using the
well-known expansion of $K_0$ for small-argument~\cite{Grad} we verify
that the small-$r$ asymptotics~(\ref{n++-beta=2-asympt}) are in
complete agreement with the exact
expression~(\ref{eq:n++-beta=2-exact}).

\section{The opposite charge case: $Q_1>0$ and $Q_2<0$}
\label{sec:charge-dislike}

If $Q_1$ and $Q_2$ have different signs, say $Q_1>0$ and $Q_2<0$, then
the dominant behavior of $E_{Q_1 Q_2}(r)$ is always given
by~(\ref{eq:small-r-dominant-n=0}), since we always have
$\beta|Q_1+Q_2|<2$, because in order to satisfy the stability
condition $0<\beta Q_1<2$ and $-2<\beta Q_2<0$. The condition for the
term $n=-1$ (or $n=+1$) to become dominant, exposed in
Sec.~\ref{sec:charge-like}, is never satisfied.  Therefore, in this
section, we will be interested in the next subdominant contribution to
$E_{Q_1 Q_2}(r)$ in some particular cases.


\subsection{Effective mean-force potential between two opposite
  external charges}
\label{sec:+Q-Q}

In this section we are interested in the case $Q_1=Q>0$ and
$Q_2=-Q$. Notice that in this case in the general small-$r$
expansion~(\ref{eq:small-r-general}), the term $n>0$ and $-n$ are of
the same order: $r^{-\beta Q^2+2n(1-\beta/4)+n^2\beta/2}$. Furthermore
besides the factor $r^{-\beta Q^2}$ which is present for all $n$, the
rest of the power of $r$ is independent of $Q$. Thus to study the
first subdominant contribution we need to take into account both terms
$n=1$ and $n=-1$. Actually, a closer look to the
expansion~(\ref{eq:small-r-general}) shows that in the limit
$Q_1\to-Q_2$, the coefficients of the expansion diverge. But the
coefficient for $n$ have a first order pole with opposite residue
than the one for $-n$, thus adding both contributions gives a finite
result. Expanding the coefficients for $n=-1$ and $n=1$
in~(\ref{eq:small-r-general}) when $Q_1+Q_2\to 0$ and adding both
contributions gives
\begin{multline}
  \label{eq:small-r+Q-Q}
  e^{-\beta E_{Q,-Q}(r)}
  \eqat{r\to0}
  \frac{r^{-\beta Q}}{\langle e^{ib Q \phi} \rangle^2}-
  \pi z r^{-\beta Q+ 2}\, 
  \frac{\langle e^{ib \phi} \rangle}{\langle e^{ib Q \phi} \rangle^2}
  \left(\frac{\beta Q}{2}\right)^2 \times
    \\
    \left[
      \frac{4}{\beta}
      I'_b(1)
      +4\ln\left[\left(\frac{\pi z}{\gamma(b^2)}\right)^{\frac{2}{4-\beta}}
	r\right]
      -4+4C+\psi(-\frac{\beta Q}{2})+\psi(\frac{\beta Q}{2})
      +\psi(1-\frac{\beta Q}{2})
      +\psi(1+\frac{\beta Q}{2})
      \right]
    \\
    +\frac{z^2}{\langle e^{ibQ\phi} \rangle^2}\,r^{-\beta Q^2+4-\beta}
    \, J(\beta Q,-\beta Q,\beta)
    +O(r^{-\beta Q^2+4})
\end{multline}
We have also written explicitly the second subdominant contribution of
order $r^{-\beta Q+4-\beta}$ which actually does not come from the
$n=-2$ and $n=2$ terms but from the first subdominant contribution to
the $n=0$ term: the last term in
equation~(\ref{eq:small-r-general}). An explicit expression for
$J(\beta Q,-\beta Q,\beta)$ is given in the appendix~\ref{app:J},
formulas~(\ref{eq:J+Q-Q}-\ref{eq:J+Q-Q-end}) .

The next subdominant contribution $O(r^{-\beta Q^2+4})$, not written
here explicitly comes from a contribution of the descendant field
$\langle (\partial\phi)^2
(\bar{\partial}\phi)^2\rangle$~\cite{Fateev-Fradkin-Lukyanov-Zad2-descendent}.


\subsection{Counterion density profile around an single external charge
  $Q>0$}
\label{sec:counterion}

The density of negative particles (counterions) of the plasma around a
single guest charge $Q>0$ is given by
\begin{equation}
  n_{-}(r)=z \frac{\langle e^{ibQ \phi(0)} e^{-ib\phi(r)}
  \rangle}{\langle e^{ibQ\phi}\rangle}
  \,.
\end{equation}
The leading term in the small-$r$ expansion is of order $r^{-\beta
Q}$, as explained earlier. Using the general
expansion~(\ref{eq:small-r-general}) we find that next subleading
contributions comes from the $n=1$ term if $Q<1$ and from the $n=-1$
term if $Q>1$. Thus,
\begin{equation}
  \label{eq:n+Q}
  n_{-}(r)
  \eqat{r\to 0}
  \begin{cases}
  z \,
  \frac{\langle e^{ib (Q-1)\phi}\rangle}{\langle e^{ibQ\phi}\rangle}
  \,r^{-\beta Q} 
  + z^2 r^{2-\beta} j_1(\beta Q,-\beta,\beta)
  +O(r^{4-\beta-\beta Q})
  & Q<1 \\
  z\,   \frac{\langle e^{ib (Q-1)\phi}\rangle}{\langle e^{ibQ\phi}\rangle}
  \,r^{-\beta Q} 
  + z^2
  \frac{\langle e^{ib(Q-2)\phi}\rangle}{\langle e^{ibQ \phi} \rangle}
  r^{-2\beta Q +\beta +2}
  j_1(-\beta Q,\beta,\beta)
  +O(r^{4-\beta-\beta Q},r^{4+3\beta-3\beta Q})
  & Q>1
  \end{cases}
\end{equation}
Notice that if $Q=1$ both subleading contributions presented above are
of the same order, in $r^{2-\beta}$, and they should be added as
explained in the preceding section. The case $Q=1$ will be studied in
the next section.

The limit $\beta Q\to 2$ is quite unnatural here. Notice that when
$\beta Q\to2$, the denominator of the leading term in $r^{-\beta Q}$
diverges. Then this term will disappear when $\beta Q=2$. On the other
hand the next subleading term, from~(\ref{eq:n+Q}), case $Q>1$, has a
finite limit when $\beta Q\to2$. Indeed, in the numerator $j_1(-\beta
Q,\beta,\beta)$ has a pole of order 1 at $\beta=2$, but so does
$\langle e^{ib Q\phi}\rangle$ in the denominator. Thus, we find
\begin{equation}
  \label{eq:n+Q-beta=2}
  n_{+}(r)\simr
    z\,
    \frac{\langle e^{ib(\frac{1}{2b^2}-2)\phi} \rangle}{\langle
    e^{ib(\frac{1}{2b^2}-1)\phi} \rangle} 
    \,r^{\beta-2} 
    \,,
    \qquad \beta Q=2
    \,.
\end{equation}

We call this limit ``unnatural'' because the power of $r$ changes in a
non-continuous manner from $-\beta Q$ to $\beta-2$, contrary to the
``natural'' limits found in Sec.~\ref{sec:charge-like}. Also the
mechanism for finding a finite result is not the same as in
Sec.~\ref{sec:charge-like}, where we added terms of the same order and
their poles canceled each other.

Let us recall the fact that at $\beta Q=2$, strictly speaking we need
to add a short-distance regularization to the potential created by the
external charge, for example by considering that the external charge
is not a point but a small impenetrable disk. With this regularization
the denominator in the leading term of~(\ref{eq:n+Q}) will not vanish
when $\beta Q=2$ and the density $n_{-}(r)$ will still behave as
$r^{-\beta Q}$, but with a prefactor which is not universal, in the
sense that it depends on the regularization procedure (it depends on
the radius of the external disk particle). This is opposed to the
situation for $n_{+}(r)$ that has the universal
limit~(\ref{eq:n+betaQ=2-beta}) when $\beta Q=2$, independent of the
regularization procedure (provided the radius of the external particle
is small enough).


\subsection{Density correlation function $n_{+-}^{(2)}(r)$}
\label{sec:n+-}

Putting $Q=1$ in equation~(\ref{eq:small-r+Q-Q}) allow us to find the
small-$r$ expansion of the density correlation function
$n_{+-}^{(2)}(r)$ between a positive charge of the plasma and a
negative one,
\begin{multline}
  \label{eq:n+-}
  n_{+-}^{(2)}(r)
  \eqat{r\to 0}
  z^{2} r^{-\beta} \Bigg\{
  1 - \pi z r^2 \frac{\beta^2}{4}\langle e^{ib \phi} \rangle
  \times \\
  \left[
    \frac{4}{\beta}I'_b(1) + 4 \ln\left[\left(\frac{\pi z}{\gamma(b^2)}
	\right)^{\frac{2}{4-\beta}}\, r\right]
    -4+4C +\psi(\frac{\beta}{2})+\psi(-\frac{\beta}{2})
    +\psi(1-\frac{\beta}{2})+\psi(1+\frac{\beta}{2})
    \right]
  \\
  +z^2 r^{4-\beta} J(\beta,-\beta,\beta)
  +O(r^{4})
    \Bigg\}
\end{multline}
An explicit expression for the coefficient $J(\beta,-\beta,\beta)$ is
given in the appendix~\ref{app:J}, see
equations~(\ref{eq:J}-\ref{eq:J-end}).

As a test, it is interesting to study the limit $\beta\to 2$ and
compare it with the exact expression from
Refs.~\cite{CornuJanco-Coulomb-gas, CornuJanco-JCP}
\begin{equation}
  \label{eq:n+-beta=2-exact}
  n_{+-}^{(2)T}(r)=(2\pi z^2) (K_1(2\pi z r))^2
  \,,
  \qquad \beta=2
\end{equation}
where $K_1(x)$ is the modified Bessel of order 1.

Notice that when $\beta\to2$ the terms in $r^{2-\beta}$ and
$r^{4-2\beta}$ from equation~(\ref{eq:n+-}) become of the same
order. Using the expansions~(\ref{eq:eibphi-beta=2}) for $\langle
e^{ib\phi}\rangle$ and~(\ref{eq:Iprimeb1-beta=2}) for $I'_b(1)$ from
appendix~\ref{app:eibQphi}, and the expansion~(\ref{eq:J-beta=2}) from
appendix~\ref{app:J} for $J(\beta,-\beta,\beta)$ when $\beta\to 2$, we
find
\begin{multline}
  n_{+-}^{(2)}(r)
  \eqat{r\to0,\,\beta\to 2}
  z^2 r^{-2}
  +(2\pi z^2)^2
  \Bigg\{
    \frac{1}{(\beta-2)^2}+\frac{2}{\beta-2}(C+\ln(\pi z))
    -\frac{1}{2}+C+\ln(\pi z r)\\
    +C+\ln(\pi z)
    +2(\ln(\pi z)+C)^2
  \Bigg\}
  +O(r^{2})+O(\beta-2)\,.
\end{multline}
Using the expansion~(\ref{eq:eibphi-beta=2-carre}) of $\langle
e^{ib\phi} \rangle^2$ we can write
\begin{equation}
  n_{+-}^{(2)}(r)
  \eqat{r\to0,\,\beta\to 2}
  z^2 r^{-2}
  +(2\pi z^2)^2
  \left[
    -\frac{1}{2}+C+\ln(\pi z r)
    \right]
  +z^2 \langle e^{ib\phi} \rangle^2
  +O(r^2) +O(\beta-2)
  \,.
\end{equation}
We recover the well-known fact that although $n^{(2)}_{+-}(r)$ does
not have a finite limit when $\beta\to2$, the truncated correlation
function $n_{+-}^{(2)T}(r)=n^{(2)}_{+-}(r)-n_{+}^b
n_{-}^b=n^{(2)}_{+-}(r)-z^2\langle e^{ib\phi}\rangle^2$ has a finite
limit when $\beta=2$,
\begin{equation}
  \label{eq:n+-beta=2-small-r}
  n_{+-}^{(2)T}(r)
  \eqat{r\to0}
  z^2 r^{-2}
  +(2\pi z^2)^2
  \left[
    -\frac{1}{2}+C+\ln(\pi z r)
    \right]+O(r^2)
  \,,\qquad \beta=2\,.  
\end{equation}
With the known expansion of $K_1(x)$ for small-argument, we verify
that the small-$r$ expansion~(\ref{eq:n+-beta=2-small-r}) is in
complete agreement with the exact
expression~(\ref{eq:n+-beta=2-exact}).


\section{Conclusion and perspectives}

Using the operator product expansion we have determined the
short-distance behavior of the effective potential between two charges
$Q_1$ and $Q_2$ immersed in a two-dimensional charge-symmetric
two-component plasma. We were also able to determine the
short-distance behavior of the density profiles around a single
external charge $Q$ in the plasma. The coion density profile exhibits
a change of behavior at $\beta Q=2-\beta$, from a dependence in
$r^{\beta}$ for $\beta Q<2-\beta$ to a dependence in $r^{2-\beta}$ for
$\beta Q>2-\beta$. On the other hand the leading behavior of the
counterion density profile is always $r^{-\beta Q}$. We have also
determined the first subleading contribution to the counterion density
profile. Specializing to the case $Q=1$, we obtained the
short-distance behavior of the density correlation functions.

This work complements those of
Refs.~\cite{Samaj-Janco-TCP-large-distance-correl,Samaj-guest-charges,
Tellez-guest-charges} in which the large-distance behavior of the
correlations and related quantities were determined. Although there is
not (yet) a closed analytical expression for the correlations
functions of the two-component plasma, we now have both its large- and
short-distance asymptotics.

A possible interesting application of this work is the following. As
we mentioned in the introduction, the bulk thermodynamic properties of
the two-component plasma have been determined for $\beta<2$ for a
system of point particles~\cite{Samaj-Travenec-TCP}. However, Kalinay
and Samaj~\cite{Kalinay-Samaj} have devised a method to obtain the
thermodynamic quantities of the two-component plasma beyond $\beta=2$
up to $\beta<3$ for a system where the charges are small disks of
radius $\sigma$, in the limit $n\sigma^2\to 0$ ($n$ is the density). A
basic ingredient of their method is the knowledge of the
short-distance behavior of the correlation functions in the region
$2<\beta<4$. Their work is based on the leading behavior, in
$r^{-\beta}$, of $n_{+-}(r)$. If the operator product expansion can be
used beyond $\beta =2$, a simple generalization of the analysis
of Secs.~\ref{sec:n++} and~\ref{sec:n+-} gives, for $2<\beta<4$,
\begin{eqnarray}
  n_{++}^{(2)}(r) 
  &\eqat{r\to0}&
  -\frac{4\pi^2 z^{4} r^{4-2\beta}}{(\beta-2)^2}
  \gamma(1-\frac{\beta}{4})^3
  \gamma(-1+\frac{3\beta}{4})
  +O(r^{2-\beta},r^{8-3\beta})
  \\
  n_{+-}^{(2)}(r)
  &\eqat{r\to0}&
  z^2 r^{-\beta}+
  z^4 r^{4-2\beta} J(\beta,-\beta,\beta)
  +O(r^{2-\beta},r^{8-3\beta})
  \,.
\end{eqnarray}
With this expansion, using the method of Ref.~\cite{Kalinay-Samaj},
the thermodynamic properties of the two-component plasma, in the
low-density limit, can be obtained up to $\beta<10/3$.


\begin{acknowledgments}
  The author acknowledge partial financial support from COLCIENCIAS
  grant 1204-05-13625,
  ECOS-Nord/COLCIENCIAS, and Comit\'e de Investigaciones de la Facultad
  de Ciencias de la Universidad de los Andes.
\end{acknowledgments}


\appendix

\section{Expansions of $\langle e^{ib Q \phi} \rangle$}
\label{app:eibQphi}

In the text we need the expansion of $\langle e^{ib Q \phi} \rangle$
around a regular value of $Q$ and also around the pole $\beta Q=2$.
If $\beta Q<2$, the expansion is immediate to obtain
\begin{equation}
  \langle e^{ib (Q + \epsilon)\phi} \rangle
  \eqat{\epsilon\to0}
  \langle e^{ib Q \phi} \rangle + 
  \epsilon \frac{\partial\langle e^{ib Q \phi} \rangle}{\partial Q}
  +\frac{\epsilon^2}{2}
  \frac{\partial^2\langle e^{ib Q \phi} \rangle}{\partial Q^2}
  +O(\epsilon^3)
  \,.
\end{equation}
Using the explicit expression~(\ref{eq:eibQ}) for $\langle e^{ib Q
\phi} \rangle$ we have
\begin{multline}
  \label{eq:eibQ+eps}
  \langle e^{ib (Q +\epsilon)\phi} \rangle
  \eqat{\epsilon\to0}
  \langle e^{ib Q \phi} \rangle
  \Bigg\{
  1+\epsilon \left[I'_b(Q)+\frac{2\beta Q}{4-\beta}
    \ln\frac{\pi z}{\gamma(b^2)}
    \right]\\
  +\frac{\epsilon^2}{2}
  \left[
    \left(I'_b(Q)+\frac{2\beta Q}{4-\beta}
    \ln\frac{\pi z}{\gamma(b^2)}
    \right)^2
    +I_b''(Q)+\frac{2\beta}{4-\beta}\ln\frac{\pi z}{\gamma(b^2)}
    \right]
    +O(\epsilon^3)
  \Bigg\}
\end{multline}
where
\begin{equation}
  \label{eq:Iprimeb}
  I'_b(Q)=\frac{\partial I_b(Q)}{\partial Q}=
  b^2 \int_0^{\infty}
  \frac{dt}{t}\,
  \left[
    -4 Q e^{-2t}
    +\frac{t \sinh(4b^2 Q t)}{\sinh t \cosh[(1-b^2)t]
      \sinh(b^2 t)}
    \right]
\end{equation}
and
\begin{equation}
  I''_b(Q)=\frac{\partial^2 I_b(Q)}{\partial Q^2}=
  4b^2\int_0^{\infty}  \frac{dt}{t}\,
  \left[
    -e^{-2t}+
    \frac{b^2 t^2 \cosh(4b^2 Qt)}{\sinh t \cosh[(1-b^2)t]
      \sinh(b^2 t)}
    \right]
  \,.
\end{equation}

If $\beta Q\to 2$, the expansion of $\langle e^{ib Q \phi} \rangle$
can be done following the same steps of the appendix of
Ref.~\cite{Samaj-guest-charges}. Let $\beta Q=2-\beta
\epsilon$. First, we isolate the pole by writing
$I_b(Q)=\I_1(\epsilon)+\I_2(\epsilon)+\I_3(\epsilon)$, with
\begin{subequations}
\begin{eqnarray}
  \I_1(\epsilon)&=&
  \int_0^{1} \frac{dt}{t}
  \left[
    \frac{\sinh^2[(1-2b^2 \epsilon)t]}{
      2\sinh(b^2t)\sinh t\sinh[(1-b^2)t]}
    -2\left(\frac{1}{2b^2}-\epsilon\right)^2 b^2 e^{-2t}
    \right]\\
  \I_2(\epsilon)&=&
  \int_{1}^{\infty} \frac{dt}{t}
  \left[
    \frac{\sinh^2[(1-2b^2 \epsilon)t]}{
      2\sinh(b^2t)\sinh t\sinh[(1-b^2)t]}
    -e^{-4b^2\epsilon t}
    -2\left(\frac{1}{2b^2}-\epsilon\right)^2 b^2 e^{-2t}
    \right]\\
  \I_3(\epsilon)&=&\int_1^{\infty} \frac{e^{-4 b^2 \epsilon t}}{t}\, dt
  =\Gamma(0,4b^2 \epsilon)\eqat{\epsilon\to0}
  -\ln(4b^2\epsilon)-C+4b^2\epsilon -4(b^2\epsilon)^2+O(\epsilon^3)
\end{eqnarray}
\end{subequations}
where $\Gamma(\alpha,x)=\int_{x}^{\infty} e^{-t} t^{\alpha-1}\,dt$ is
the incomplete Gamma function. The integral $\I_3$ contains the pole
at $\epsilon=0$, while the integrals $\I_1$ and $\I_2$ are regular at
$\epsilon=0$. The expansion of $\I_1$ and $\I_2$ can simply be
obtained with their Taylor series around $\epsilon=0$:
$\I_{1,2}(\epsilon)=\I_{1,2}(0)+\epsilon\,
\I_{1,2}'(0)+(\epsilon^2/2)\I_{1,2}''(0)+O(\epsilon^3)$. 

Putting together all the contributions from $\I_1$, $\I_2$, $\I_3$ and
from the expansion of $(\pi z/\gamma(b^2))^{Q^2 \xi}$
in~(\ref{eq:eibQ}), we finally find
\begin{multline}
  \label{eq:eibQ-betaQ=2}
  \langle e^{ib Q \phi} \rangle
  \eqat{\beta Q\to 2}
  2\pi z \langle e^{ib(\frac{1}{2b^2}-1)\phi} \rangle
  \Bigg\{
  \frac{1}{2-\beta Q}
  +\frac{R_1(\beta/4)}{\beta}-\frac{4}{\beta(4-\beta)}
  \ln\frac{\pi z}{\gamma(\beta/4)}
  \\
  +
  \frac{2-\beta Q}{\beta^2}
  \left[
    R_2(\beta/4)+\frac{\beta}{4-\beta}\ln\frac{\pi z}{\gamma(\beta/4)}
    +\frac{1}{2}\left(R_1(\beta/4)-\frac{4}{4-\beta}\ln\frac{\pi
  z}{\gamma(\beta/4)} \right)^2
    \right]
  +O((2-\beta Q)^2)
  \Bigg\}
\end{multline}
with
\begin{eqnarray}
  \label{eq:R1}
  R_1(b^2)&=&
  2\int_0^{\infty}
  \frac{dt}{t}\,
  \left[
    e^{-2t}+
    tb^2\left(
    2-\frac{\cosh t}{\sinh(b^2 t)\cosh[(1-b^2)t]}
    \right)
    \right]
  \\
  R_2(b^2)&=&2b^2\int_0^{\infty}
  \frac{dt}{t}\left[
    b^2t^2\left(
    \frac{\cosh t \coth t +\sinh t}{\sinh(b^2 t) \cosh[(1-b^2)t]}
    -4
    \right)
    -e^{-2t}
    \right]
  \,.
\end{eqnarray}
Actually, in the main text, we only use the expansion up to order
$O(1)$, the $O(\beta Q-2)$ term is given here only for informational
purposes.

It is also useful to know the expansion of $\langle e^{ib\phi}\rangle$
when $\beta\to2$. From the explicit formula~(\ref{eq:eibphi}) we find
\begin{equation}
  \label{eq:eibphi-beta=2}
  \langle e^{ib\phi}\rangle
  \eqat{\beta\to2}
  \frac{2\pi z}{2-\beta}
  -2\pi z [C+\ln(\pi z)]
  -\pi z (\beta-2)
  [C+\ln(\pi z)][1+C+\ln(\pi z)]
  +O((\beta-2)^2)
\end{equation}
and
\begin{equation}
  \label{eq:eibphi-beta=2-carre}
  \langle e^{ib\phi}\rangle^2
  \eqat{\beta\to2}
  (2\pi z)^2
  \left[
    \frac{1}{(\beta-2)^2}+\frac{2}{\beta-2}
    [C+\ln(\pi z)]
    +C+\ln(\pi z)+2[C+\ln(\pi z)]^2
    +O(\beta-2)
    \right]
  \,.
\end{equation}

In Sec.~\ref{sec:n+-} we need the expansion of $I'_b(1)$ when
$\beta\to 2$ which is a pole of first order. To isolate the pole
contribution, it is convenient to write $I_b'(1)$ as
\begin{equation}
  \frac{4}{\beta}I'_b(1)=
  \frac{4}{2-\beta}+
  \int_0^{\infty}
  \left[-\frac{4e^{-2t}}{t}+
  \frac{\sinh(\beta t)}{\sinh t\sinh(\beta
  t/4)\cosh[(1-\frac{\beta}{4})t]}
  -4 e^{-(2-\beta) t}
  \right]
  \,dt
  \,.
\end{equation}
The remaining integral is now convergent when $\beta\to2$, and can be
easily expanded around $\beta=2$. We find
\begin{equation}
  \label{eq:Iprimeb1-beta=2}
  \frac{4}{\beta}I'_b(1)=
  \frac{4}{2-\beta}
  +4C -4(\beta-2)\ln 2+O((\beta-2)^2)
  \,.
\end{equation}


\section{Computation of $J(\beta Q,-\beta Q,\beta)$}
\label{app:J}

In this appendix we give an explicit expression for the integral
\begin{equation}
  \label{eq:Dotsenko-integral-J-part}
  J(\beta Q,-\beta Q,\beta)=
  \int d^2 x \,d^2 y\,
  \frac{|x|^{\beta Q} |1-y|^{\beta Q}}{|y|^{\beta Q}
  |1-x|^{\beta Q} |x-y|^{\beta}}
  \,.
\end{equation}
In the appendix D of Ref.~\cite{Dotsenko-Picco-Pujol}, an integral
more general than~(\ref{eq:Dotsenko-integral-J-part}), with arbitrary
powers in each factor is computed. Using the general results from
Ref.~\cite{Dotsenko-Picco-Pujol}, we find for our particular case that
\begin{equation}
  \label{eq:J+Q-Q}
  J(\beta Q,-\beta Q,\beta)=
  s(\beta/2)^2 \left\{
  (J^{-}_{1})^2 + (J_{2}^{-})^2
  \right\}
\end{equation}
where
\begin{eqnarray}
  J_{1}^{-}&=&-
  \frac{s(\beta Q/2)}{s(\beta/2)^2}
  \left(
  s(\beta Q/2) J_{1}^{+}
  +s(\beta(Q+1)/2) J_{2}^{+}
  \right)\\
  J_{2}^{-}&=&-
  \frac{s(\beta Q/2)}{s(\beta/2)^2}
  \left(
  s(\beta(Q-1)/2) J_{1}^{+}+
  s(\beta Q/2) J_2^{+} 
  \right)
\end{eqnarray}
where we used the notation $s(x)=\sin (\pi x)$ and
\begin{eqnarray}
  J_{1}^{+}&=&
  \frac{\Gamma(1-\frac{\beta}{2})\Gamma(2-\frac{\beta}{2})
    \Gamma(1-\frac{\beta Q}{2})^2}{
    \Gamma(2-\frac{\beta(Q+1)}{2})\Gamma(3-\frac{\beta(Q+1)}{2})}
  \times\nonumber\\
    &&\hspace{1cm}
  \F32 \left(2-\frac{\beta}{2},-\frac{\beta Q}{2},1-\frac{\beta Q}{2};
    2-\frac{\beta(Q+1)}{2}, 3-\frac{\beta(Q+1)}{2};1\right)\\
    J_{2}^{+}&=&
    \frac{\Gamma(1-\frac{\beta}{2})\Gamma(2-\frac{\beta}{2})
      \Gamma(1+\frac{\beta Q}{2})^2}{
      \Gamma(2+\frac{\beta(Q-1)}{2})
      \Gamma(3+\frac{\beta(Q-1)}{2})}
    \times\nonumber\\
    &&\hspace{1cm}
    \F32 \left(2-\frac{\beta}{2},\frac{\beta Q}{2},1+\frac{\beta Q}{2};
    2+\frac{\beta(Q-1)}{2}, 3+\frac{\beta(Q-1)}{2}; 1\right)
    \label{eq:J+Q-Q-end}
\end{eqnarray}
where $\F32(a_1,a_2,a_3;b_1,b_2;z)=\sum_{k=0}^{\infty}
\frac{(a_1)_k(a_2)_k (a_3)_k}{(b_1)_k (b_2)_k k!}z^k$ is a
generalized hypergeometric function, and
$(a)_k=\Gamma(a+k)/\Gamma(a)$ is the Pochhammer symbol.

For the particular case $Q=1$, the above expressions are simplified:
\begin{equation}
  \label{eq:J}
  J(\beta,-\beta,\beta)=
  \left[
    s(\beta/2) J_{1}^{+}+s(\beta)J_{2}^{+}
    \right]^2+
  \left[
    s(\beta/2) J_{2}^{+}
    \right]^2
\end{equation}
with
\begin{eqnarray}
  J_{1}^{+}&=&
  \frac{\Gamma(1-\frac{\beta}{2})^3\Gamma(2-\frac{\beta}{2})}{
    \Gamma(2-\beta)\Gamma(3-\beta)}\,
  \F32 \left(1-\frac{\beta}{2},2-\frac{\beta}{2},-\frac{\beta}{2};
  2-\beta,3-\beta;1\right)\\
  J_{2}^{+}&=&
  \frac{\Gamma(1-\frac{\beta}{2})\Gamma(2-\frac{\beta}{2})
    \Gamma(1+\frac{\beta}{2})^2
  }{2}\,
  \F32 \left(2-\frac{\beta}{2},1+\frac{\beta}{2},\frac{\beta}{2};
  2,3;1\right)\,.
  \label{eq:J-end}
\end{eqnarray}

To study the limit $\beta\to 2$, we need the expansion of
$J(\beta,-\beta,\beta)$ around $\beta=2$. In the hypergeometric
function defining $J_{2}^{+}$, $\beta=2$ is a regular point. We can
obtain its expansion with the Taylor series
\begin{multline}
\F32\left(2-\frac{\beta}{2},1+\frac{\beta}{2},\frac{\beta}{2}; 2,3;1\right)
\eqat{\beta\to2}
\F32(2,1,0; 2,3;1) 
\\+ (\beta-2)\left.\frac{\partial
\F32(2-\frac{\beta}{2},1+\frac{\beta}{2},\frac{\beta}{2};
2,3;1)}{\partial \beta}\right|_{\beta=2}+O((\beta-2)^2)
\,.  
\end{multline}
Deriving under the summation we find
\begin{equation}
\left.\frac{\partial
\F32(2-\frac{\beta}{2},1+\frac{\beta}{2},\frac{\beta}{2};
2,3;1)}{\partial \beta}\right|_{\beta=2}
=
\sum_{k=1}^{\infty}\frac{-1+C+\psi(2+k)}{(k+1)(k+2)}
= \frac{\pi^2}{6}-1
\,.
\end{equation}
Then
\begin{equation}
\F32 \left(2-\frac{\beta}{2},1+\frac{\beta}{2},\frac{\beta}{2}; 2,3;1\right)
\eqat{\beta\to2}
2+(\beta-2)\left(\frac{\pi^2}{6}-1\right)+O((\beta-2)^2)
\,.
\end{equation}
With this result we have the following expansions, needed for the
expansion of $J(\beta,-\beta,\beta)$,
\begin{eqnarray}
  \label{eq:J2a}
  s(\beta/2) J_{2}^{+}
  &\eqat{\beta\to2}&
  \pi + \frac{\pi}{12}\left(6+\pi^2\right)(\beta-2)
  +O((\beta-2)^2)\\
  \label{eq:J2b}
  s(\beta)J_{2}^{+}
  &\eqat{\beta\to2}&
  -2\pi-\frac{\pi}{6}\left(6+\pi^2\right)(\beta-2)
  +O((\beta-2)^2)
\end{eqnarray}
For the expansion of the hypergeometric function in $J_{1}^{+}$, it is
convenient to write it as
\begin{equation}
  \F32 \left(1-\frac{\beta}{2},2-\frac{\beta}{2},-\frac{\beta}{2};
  2-\beta,3-\beta;1\right)=
  1-\frac{\frac{\beta}{2}(2-\frac{\beta}{2})}{2(3-\beta)}
  +(1-\frac{\beta}{2})S(\beta)  
\end{equation}
with
\begin{equation}
  S(\beta)=-\sum_{k=2}^{\infty}
  \frac{(\beta/2)(2-\frac{\beta}{2})_{k-2}
    (2-\frac{\beta}{2})_{k-1}(2-\frac{\beta}{2})_{k}}{
    2 (3-\beta)_{k-1} (3-\beta)_{k}\, k!}
\,.
\end{equation}
This last term can be expanded to the first order using
$S(\beta)\eqat{\beta\to2}S(2)+(\beta-2)S'(2)+O(\beta-2)$, with
$S'(2)=-\sum_{k=2}^{\infty}
\frac{1+C-\psi(k-1)+\psi(k)+\psi(k+1)}{4k(k-1)}=-(1+\pi^2/12)/2$. Then,
\begin{equation}
  S(\beta)\eqat{\beta\to2}-\frac{1}{2}-\frac{\beta-2}{2}
  \left(1+\frac{\pi^2}{12}\right)
  +O((\beta-2)^2)
\end{equation}
and finally
\begin{equation}
  \label{eq:F32-J1}
  \F32 \left(1-\frac{\beta}{2},2-\frac{\beta}{2},-\frac{\beta}{2};
  2-\beta,3-\beta;1\right)\eqat{\beta\to2}
  \frac{1}{2}-\frac{\beta-2}{4}
  +\frac{(\beta-2)^2}{8}
  \left(
  \frac{\pi^2}{6}-1\right)
  +O((\beta-2)^3)
  \,.
\end{equation}
We expanded up to order $(\beta-2)^2$, since $J_{1}^{+}$ has a pole of
order 2 at $\beta=2$, as opposed to $J_{2}^{+}$ which as a pole of
order 1. With this expansion~(\ref{eq:F32-J1}), we find
\begin{equation}
  \label{eq:J1}
  s(\beta/2) J_{1}^{+}
  \eqat{\beta\to2}
  -\frac{2\pi}{\beta-2}
  +\pi
  +\frac{\pi}{6}(3+\pi^2)(\beta-2)
  +O((\beta-2)^2)
  \,.
\end{equation}
Putting together equations~(\ref{eq:J2a}), (\ref{eq:J2b}),
and~(\ref{eq:J1}) we finally find
\begin{equation}
  \label{eq:J-beta=2}
  J(\beta,-\beta,\beta)
  \eqat{\beta\to2}
  (2\pi)^2 \left[
    \frac{1}{(\beta-2)^2}+\frac{1}{\beta-2}
    +1
    \right]
  +O(\beta-2)
  \,.
\end{equation}


\end{document}